\documentclass[12pt]{article}

\usepackage{graphicx}
\usepackage{lscape}

  \def\be{\begin{equation}}      
  \def\ee{\end{equation}}    \def\beq{\begin{eqnarray}}
      \def\eeq{\end{eqnarray}}
       \def\m{\multicolumn}

\begin{document}

\begin{center}
{\large \textbf{Mixing and lifetime of B-meson using Coulomb plus power potential}}\\
\textbf{Bhavin Patel$^1$, Ajay Kumar Rai$^2$ and P.C.Vinodkumar$^1$\\}

 $^1$Department of Physics, Sardar Patel University,\\
  Vallabh Vidyanagar- 388 120, Gujarat, INDIA\\
$^2$Physics Section, Department of Applied Sciences \& Humanities,\\
Sardar Vallabhbhai National Institute of Technology, Surat-395 007,
Gujarat, INDIA.
\end{center}

\abstract{
The investigation of mixing phenomena and lifetime in neutral B
meson systems provides an important testing ground for standard
model flavour dynamics. Spectroscopic parameters has been used to
calculate the pseudoscalar decay constant and predictions of mixing
mass parameters and lifetimes of the B$_{d}$ and  B$_{s}$ mesons.
}
\section{Introduction}
The study of mixing and lifetimes of the $B$-meson, provides us
useful information about the dynamics of quarks and gluons at the
hadronic scale. The remarkable progress at the experimental side,
with various high energy machines such as LHC, B-factories,
Tevatron, ARGUS collaborations, CLEO, CDF, DO \emph{etc.} for the
study of hadrons has opened up new challenges in the theoretical
understanding of B-meson. In order to understand the structure of
the newly observed zoo of open flavour meson
resonances\cite{Aubert2007,CLEO2003,PDG2006,Antimo2006} in the
energy range of 2-5 Ge$V$, it is necessary to analyze their
spectroscopic properties and  decay modes based on theoretical
models. Many of these states could be the excited beauty mesonic
states while for many other states the possibility of multi-quark or
molecular like structures are being proposed. Thus, the main
objective of the present paper includes the study of decay constant,
mixing parameters and lifetimes of the open flavour beauty mesons
using the spectroscopic parameters. We study these open beauty
mesons within the frame work of a potential model
\cite{AKRai2002,AKRai2005}. The weak decay constants of pseudoscalar
and vector decay constants are important in the calculations of
various decay rates \cite{Ebert2006} and mixing parameters of the
mesons.\\
\begin{figure}[h]
\begin{center}
\includegraphics[height=3.0in,width=3.5in]{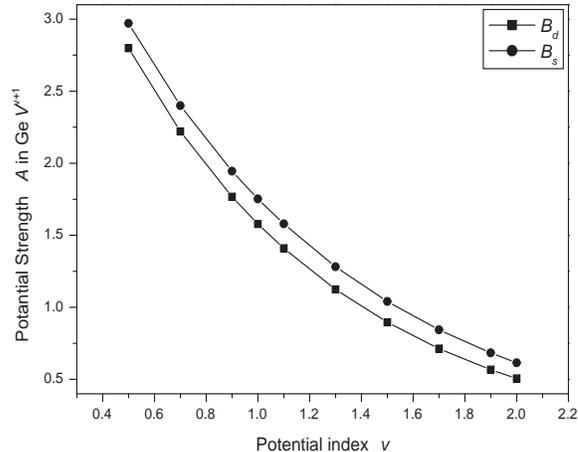}
\vspace{-0.2in} \vspace{-0.2in} \caption{Behavior of $A$ with the
potential index $\nu$ for $B_d$ and $B_s$ mesons.
Ge$V$}\label{fig:R1s}
\end{center}
\end{figure}
\section{Theoretical methodology: A Potential Scheme}
In the limit of heavy quark mass $m_Q$ $\rightarrow$ $\infty$, heavy
meson properties are governed by the dynamics of the light quark. As
such, these states become hydrogen like atom of hadron physics. We
investigate the heavy-light mass spectrum of $B$ and $B_{s}$ mesons
 in the frame work of a relativistic potential
model. For the light-heavy flavour bound system of $q\bar{Q}$ or
$\bar{q}Q$, we treat the heavy-quark $(Q = b)$ as well as the
light-quark $(q =  d, s)$ relativistically within a potential bound
mesonic system. The Hamiltonian for the case be written as
\begin{equation} \label{eq:hamiltonian}
H=\sqrt{p^2+m^{2}_{q}}+\sqrt{p^2+m^{2}_{Q}}+V(r)
%+V_{S_{\bar {Q}} \cdotS_q}(r) + V_{{L} \cdot S}(r)
\end{equation}
Where $m_{q}$ and $m_{Q}$ are the light and heavy quark mass
respectively, $p$ is the relative momentum of each quark, $V(r)$ is
the confined part of the quark-antiquark potential. Here, we
consider a general power potential with a colour Coulomb term of the
form \cite{AKRai2002,Bhavin2008}
 \begin{equation}
 V(r)=\frac{-\alpha_c}{r}+A r^{\nu}
 \end{equation}
where $\alpha_c=\frac {4}{3}\alpha_s$, $\alpha_s$ being the strong
running coupling constant, $A$ and $\nu$ are the  potential
parameters. We employ the trial wave function,
\begin{equation} \label{eq:hydrogenic}
R(r)= 2 \sqrt{\frac{\mu^{3}}{\sqrt{\pi}}}\ \  e^{-\mu^{2} r^{2}/2}
\end{equation} and use the virial
theorem, to get spin average mass from the hamiltonian defined by
Eqn.(\ref{eq:hamiltonian}). Here $\mu$ is the variational parameter.
 The parameters used here are
$m_{u/d}=0.36\ GeV$, $m_{s}=0.4 \ GeV$, $m_{b}=4.2 \ GeV$, and
$\alpha_c=0.36$ (for $B$ meson) and $\alpha_c=0.33$ (for $B_{s}$
meson). The remanning model parameter $A$ is fixed for each choices
of $\nu$ so as to get the experimental ground state spin average
masses of $B$ and $B_s$ mesons. The spin average masses of $B-B^{*}$
and the $B_{s}-B_{s}^{*}$ mesons are computed using the experimental
values of $M_B=$ 5.279 Ge$V$, $M_{B^*}=$ 5.325 Ge$V$, $M_{B_s}=$
5.369 Ge$V$, $M_{B_s^*}=$ 5.417 Ge$V$ respectively \cite{PDG2006}.
The behavior of potential strength $A$ with power index $\nu$ is
shown in Fig \ref{fig:R1s}. It is found to have an exponential
behavior with $\nu$. For computing the mass difference between
$B-B^{*}$ and the $B_{s}-B_{s}^{*}$ mesonic states, we consider the
spin dependent part of the usual OGEP given by
 \cite{Voloshin2008}
\begin{equation} \label{SS}
V_{SS}(r)=\frac{4}{3}\frac{ \ \pi \alpha_{c}}{\ m_{Q} m_{\bar q} }
\left[S(S+1)-\frac{3}{2}\right] \delta^{(3)}(\vec{r})
\end{equation}
The computed results of S-state of $B$ and $B_{s}$ with different
choices of $\nu=0.5$ to 2.0 are listed in Table-\ref{tab:1} along
the existing experimental as well as with other theoretical model
predictions.
\begin{table*}
\caption{Mass of $B_{q} (q \ \epsilon \ d, s)$ mesons} \label{tab:1}
\begin{tabular}{lllllllc}
\hline
\m{0}{c}{Meson}&\m{0}{c}{$\nu$}&\m{0}{c}{$\mu$}&\m{0}{c}{$|R(0)|^2$}&\m{2}{c}
{\underline{$M_P$ in Ge$V$}} &\m{2}{c} {\underline{$M_V$ in
Ge$V$}}\\
&&Ge$V$&Ge$V^{3}$&\m{0}{c}{Our}&\m{0}{c}{Others}&\m{0}{c}{Our}&\m{0}{c}{Others} \\
\hline
$B_d$   &   0.5 &   0.450   &   0.206   &   5.291   &       &   5.324   &           \\
    &   0.7 &   0.472   &   0.237   &   5.287   &       &   5.325   &           \\
    &   0.9 &   0.487   &   0.261   &   5.285   &       &   5.327   &           \\
    &   1.0 &   0.493   &   0.271   &   5.284   &   5.279\cite{PDG2006} &   5.327   &   5.324\cite{PDG2006}         \\
    &   1.1 &   0.499   &   0.280   &   5.283   &   5.279\cite{Pierro2001}  &   5.327   &   5.324\cite{Pierro2001}      \\
    &   1.3 &   0.508   &   0.295   &   5.282   &   5.285\cite{Ebert2003}   &   5.329   &   5.324\cite{Ebert2003}       \\
    &   1.5 &   0.515   &   0.308   &   5.279   &       &   5.328   &           \\
    &   1.7 &   0.522   &   0.320   &   5.278   &       &   5.329   &           \\
    &   1.9 &   0.527   &   0.330   &   5.277   &       &   5.329   &           \\
    &   2.0 &   0.529   &   0.334   &   5.276   &       &   5.329   &           \\
    &       &       &       &       &       &       &           \\
$B_s$   &   0.5 &   0.491   &   0.267   &   5.378   &       &   5.413   &           \\
    &   0.7 &   0.515   &   0.308   &   5.374   &       &   5.414   &           \\
    &   0.9 &   0.532   &   0.339   &   5.371   &       &   5.415   &           \\
    &   1.0 &   0.539   &   0.353   &   5.370   &   5.369\cite{PDG2006}     &   5.416   &   5.416\cite{PDG2006}         \\
    &   1.1 &   0.545   &   0.365   &   5.369   &   5.373\cite{Pierro2001}  &   5.416   &   5.421\cite{Pierro2001}      \\
    &   1.3 &   0.555   &   0.385   &   5.365   &   5.375\cite{Ebert2003}   &   5.416   &   5.412\cite{Ebert2003}       \\
    &   1.5 &   0.563   &   0.403   &   5.365   &       &   5.418   &           \\
    &   1.7 &   0.570   &   0.418   &   5.364   &       &   5.418   &           \\
    &   1.9 &   0.576   &   0.431   &   5.363   &       &   5.419   &           \\
    &   2.0 &   0.578   &   0.436   &   5.362   &       &   5.419   &           \\

\hline
\end{tabular}
\end{table*}
\begin{table*}
\caption{Decay constant,Mixing Mass Parameter and lifetime of $B_{q}
(q \ \epsilon \ d, s)$ mesons} \label{tab:2}
\begin{tabular}{lllllllc}
\hline \m{0}{c}{Meson}&\m{0}{c}{$\nu$}&\m{2}{c} {\underline{$f_P$ in
Me$V$}}
&\m{2}{c}{\underline{$\Delta m_{q}$ in $ps^{-1}$}}&\m{2}{c} {\underline{Lifetime $\tau$ in $ps$}}\\
&&\m{0}{c}{Our}&\m{0}{c}{Others}&
\m{0}{c}{Our}&\m{0}{c}{Others}&\m{0}{c}{Our}&\m{0}{c}{Others}  \\
\hline
$B_d$   &   0.5 &   192   &       &   0.45    &       &   1.55    &       \\
    &   0.7 &   206   &       &   0.52    &       &   1.56    &       \\
    &   0.9 &   216   &       &   0.57    &       &   1.56    &       \\
    &   1.0 &   220   &   196(29)\cite{Gvetic2004}    &   0.59    &   0.50\cite{PDG2006}  &   1.57    &   1.530   \\
    &   1.1 &   223   &   189 \cite{Ebert2006}    &   0.61    &       &   1.57    &   $\pm$0.009\cite{PDG2006}    \\
    &   1.3 &   230   &   190(7)$^{+24}_{-1}$\cite{Bernard2002}   &   0.65    &       &   1.57    &       \\
    &   1.5 &   235   &   203(23) \cite{Narison2001}      &   0.68    &       &   1.57    &       \\
    &   1.7 &   239   &   206(20)\cite{Penin2002} &   0.70    &       &   1.58    &       \\
    &   1.9 &   243   &   210 (9)\cite{Jamin2002} &   0.72    &       &   1.58    &       \\
    &   2.0 &   244   &       &   0.73    &       &   1.58    &       \\
    &       &       &       &       &       &       &       \\
$B_s$   &   0.5 &   217   &       &   17.67   &       &   1.43    &       \\
    &   0.7 &   233   &       &   20.35   &       &   1.43    &       \\
    &   0.9 &   244   &       &   22.40   &       &   1.44    &       \\
    &   1.0 &   249   &   216\cite{Gvetic2004}    &   23.26   &   17\cite{PDG2006}    &   1.44    &   1.466   \\
    &   1.1 &   253   &   218 \cite{Ebert2006}        &   24.04   &       &   1.44    &   $\pm$0.059\cite{PDG2006}    \\
    &   1.3 &   260   &   217(6)$^{+32}_{-28}$\cite{Bernard2002}  &   25.38   &       &   1.45    &       \\
    &   1.5 &   266   &   236(30) \cite{Narison2001}  &   26.52   &       &   1.45    &       \\
    &   1.7 &   271   &   236$\pm$30\cite{Penin2002}  &   27.49   &       &   1.45    &       \\
    &   1.9 &   275   &    244 (21)\cite{Jamin2002}   &   28.33   &       &   1.45    &       \\
    &   2.0 &   277   &       &   28.71   &       &   1.45    &       \\

\hline
\end{tabular}
\end{table*}

\section{The Decay constants of the neutral open beauty mesons}
The decay constants of mesons are important parameters in the study
of leptonic or non-leptonic weak decay processes and in the neutral
$B- \bar B$ mixing process.  In the nonrelativistic limit, the decay
constant can be expressed through  the ground state wave function at
the origin $\psi_{P}(0)$ by the Van-Royen-Weisskopf formula
\cite{Vanroyenaweissskopf}. Though most of the models predict the
mesonic mass spectrum successfully, there are disagreements in the
predictions of their  decay constants \cite{Hwang1997}.  So, we
reexamine the predictions of the decay constants with different
choices of potential index $\nu$. We consider the nonrelativistic
expression for $f_p$ as \cite{Vanroyenaweissskopf}
\begin{equation}
f^2_{P}=\frac{3 \left| R_{p}(0)\right|^2} { \pi M_{P} }
\label{eq:414}
\end{equation}
 The results  computed for $B$ and $B_{s}$ mesons
are tabulated in Table-\ref{tab:2} along with other theoretical
model predictions.
\section{Mixing Mass Parameter of $B_{q} (q \ \epsilon \ d, s)$ mesons }
In the standard model, the transitions $B^{0}_{q}-\bar{B}^{0}_{q}$
and $\bar{B}^{0}_{q}-{B}^{0}_{q}$ are due to the weak interaction.
The netural $B_{d}$ and $B_{s}$ mesons mix with their antiparticles
leading to oscillations between the mass eigenstates. This mass
oscillation is parameterized as the mixing mass parameter given by
\cite{Buras2003}
\begin{equation}
\Delta m_{B}=\frac{G_{F}^{2} m_{t}^{2} M_{B_{q}} f_{B_{q}}^{2}}{6
\pi^{2}} g(x_{t}) \eta_{t} |V_{tq}^{*}V_{tb}|^{2} B
\end{equation}
where $\eta_{t}\approx 0.55$ is the glunoic correction to the
oscillation \cite{Buras1990}, $B$ is the bag parameter and its value
is taken from the lattice result as 1.34 \cite{Buras2003}, while the
pseudoscalar mass ($M_{B_{q}}$) and the pseudoscalar decay constant
($f_{B_{q}}$) of the beauty mesons are taken as our spectroscopic
parameters determined using potential models. The $g(x_{t})$ factor
is the Inami-Lim function  given by
\begin{equation}
g(x_{t})=\frac{1}{4}+\frac{9}{4(1-x_{t})}-\frac{3}{2(1-x_{t})^{2}}-\frac{3}{2}\frac{x_{t}^{2}}{(1-x_{t})^{3}}
ln x_{t}
\end{equation}
Here, $x_{t}=\frac{m_{t}^{2}}{M_{W}^{2}}$. The values of $m_{t}$
(174 Ge$V$), $M_{W}$ (80.403 Ge$V$) and the  CKM matrix elements
$V_{tb} (1)$, $V_{td} (7.4\times10^{-3})$ and $V_{ts}
(40.6\times10^{-3})$ are taken from the Particle Data Group
\cite{PDG2006}. The results are also tabulated in Table-\ref{tab:2}.
\section{The Lifetime of the neutral open beauty mesons}
 We compute the lifetime of these
neutral $B_q (q \ \epsilon \ d, s)$ mesons by computing their
semileptonic decay widths and by using their experimental branching
ratios. The inclusive semileptonic decay width of the open beauty
flavour mesons are computed using the expression given by
 \begin{equation} \label{eq:pertqcdcorr}
\Gamma(B  \rightarrow l \bar \nu_{l} X_c)= \frac{G_{F}^{2}
m_{b}^{5}}{192\ \pi^{3}} |V_{cb}|^2 \
[f\left({x}\right)-\frac{\alpha_s}{\pi} g (x )]
\end{equation}
Here, $m_b$ the mass of $b$ quark. Generally, it is taken as the
model mass parameter coming from the  fitting  of its mass spectrum
. Within the potential confinement scheme, we consider it as the
effective mass of the quarks, $m_{q}^{eff}$ \cite{Close1982}. Thus,
effective $b$ quark mass ($m_{b}^{eff}$) is defined as
\begin{equation}\label{eq:417} m^{eff}_{b}=m_b\left(
1+\frac{E_{bind}}{m_{b}+ m_{\bar{q}}}\right) \\
\end{equation}
to account for its bound state effects. The binding effect has been
calculated as $E_{bind}=M_{bq}-(m_{b}-m_{q})$,  where $m_b$ and
$m_{q}$ are the model mass parameters employed in its spectroscopic
study and $M_{bq}$ is the mass of the mesonic state. The effective
mass of the quarks would be different from the adhoc choices of the
model mass parameters. For example, within the meson the mass of the
quarks may get modified due to its binding interactions with other
quark. Thus, the effective mass of the  $b$ quark will be different
when it is in $b\bar{s}$ combinations or in $b\bar{d}$ combinations
due to the residual strong interaction effects of the bound systems.
 The functions $f(x)$ and $g(x)$ appeared in Eqn. \ref{eq:pertqcdcorr}
 correspond to the phase space correction and the QCD correction
 at the $bc$ vertex in this decay. They are computed
 from \cite{Yosef1989,Cabibbo1978}, where the parameter  $x$ is computed as $x=
\left(\frac{m_c}{m_{b}^{eff}}\right)^2$. The lifetime of $B_d$ and
$B_s$ mesons are computed from the relation $BR(B_{q}  \rightarrow l
\bar \nu_{l} X_c)= \Gamma(B_{q}  \rightarrow l \bar \nu_{l} X_c)
\tau_{B_{q}}$. The results are listed in Table-\ref{tab:1} with
other theoretical and experimental values for comparisons.
\section{Results and Discussion}
We have employed the coulomb plus power potential form to study the
the mixing mass parameters and lifetimes of the $B_{d}$ and $B_{s}$
mesons using the spectroscopic ground state parameters. Here we
solved the Schr\"{o}dinger equation using variational approach
\cite{AKRai2002,AKRai2005}. Our potential parameters are fixed with
respect to the center of weight ground state $1S$ mass of the
$B_{q}$ mesons.  In Table-\ref{tab:2}, we tabulate our predictions
for the $f_{p}$ and $\Delta m_{q}$ and lifetime of the neutral open
beauty mesons. Our results are  compared with the available
experimental / other theoretical values. The prediction for $\Delta
m_{q}$ and lifetime of the $B_q$ mesons are very close to the
experimentally observed values. It can be seen that the mixing mass
parameter is more sensitive to the choice of the potential index
while the life time is least sensitive to the potential choice. In
conclusion we find the overall predictions in accordance with the
experimental values in the range of potential index
$0.5\leq\nu\leq0.7$. Thus the present study is an attempt to
indicate the importance of spectroscopic parameters (\emph{i.e. }
masses and decay constants) in the
weak decay processes.\\
{\bf Acknowledgement:} Part of this work is done with a financial
support from DST, Government of India, under a Major Research
Project \textbf{SR/S2/HEP-20/2006}.

\end{document}